\documentstyle[12pt,fleqn,epsfig]{article}
\newcommand{\e}{\epsilon}
\begin{document}

\begin{titlepage}
\setcounter{page}{1}

\begin{flushright}
{\bf MC-TH-95/15}\\
{\bf LAL-95/55}\\
{\bf arch-ive/9509307 }\\
{August 1995}
\end{flushright}

\vspace{1cm}

\begin{center}
\begin{Huge}
{\bf SVD Approach to Data Unfolding\\}
\end{Huge}
\vspace{5mm}
\begin{large}
Andreas H\"ocker\\
\end{large}
\bigskip
{\em Laboratoire de l'Acc\'el\'erateur Lin\'eaire,\\
IN2P3-CNRS et Universit\'e de Paris-Sud, \\
F-91405 Orsay. \\
E-mail: hoecker@lalcls.in2p3.fr\\}
\bigskip
{\em and\\}
\bigskip
\begin{large}
{Vakhtang Kartvelishvili$~{}^{a)}$}\\
\end{large}
\bigskip
{\em Department of Physics and Astronomy,\\
Schuster Laboratory, University of Manchester,\\
Manchester M13 9PL, U.K.\\
E-mail: vato@a3.ph.man.ac.uk\\}

\vspace{1cm}

{\bf Abstract}
\end{center}

\vspace{0.1cm}

\noindent
Distributions measured in high energy physics experiments are
usually distorted and/or transformed by various detector effects.
A regularization method for unfolding these distributions
is re-formulated in terms of the Singular Value Decomposition (SVD)
of the response matrix. A relatively simple, yet quite
efficient unfolding procedure is explained in detail.
The concise linear algorithm results in a straightforward
implementation with full error propagation,
including the complete covariance matrix and its inverse.
Several improvements upon widely used procedures are
proposed, and recommendations are given how to simplify the task by
the proper choice of the matrix. Ways of determining the optimal
value of the regularization parameter are suggested and discussed,
and several examples illustrating the use of the method are presented.

\vspace{25mm}

\noindent{---------------------}\\
${}^{a)}$ On leave from High Energy Physics Institute, Tbilisi State
University, Tbilisi, GE-380086, Republic of Georgia.

\end{titlepage}

\setcounter{page}{1}

\section{Introduction}

In high energy physics, measurements of physical observables ---
spectra of invariant masses, angular distributions of particles, etc. ---
are usually distorted and transformed by various effects such as
finite resolution and limited acceptance of the detector.
For this reason, it is often impossible, or at least very difficult,
 to make direct comparisons of the
data obtained using different detectors with each other and with
various theoretical predictions.

In order to overcome this problem, some sophisticated studies of
the measurement process are usually carried out. Ideally, these studies
should result in a two-variable function describing
the response of the detector, so that the actual measured distribution
can be considered as a convolution of this function with the true
one. This in general leads to an integral equation
for the true distribution. Solving this equation
(i.e. {\it{unfolding}} the true distribution) usually requires
some kind of discretization, leading to a system of linear equations.
The problem, however, belongs to a class of ill-posed problems,
which are unstable against small variations in the initial system.
Because of inevitable statistical errors in the measured distribution,
 the exact solution (if it exists) is usually wildly oscillating and
useless (see ref. \cite{Blobel} for a good introduction to the subject
{\footnote{Many of the relevant topics are compiled also in the
recent publication \cite{Zech}.}}
).

For a number of reasons, in present high energy physics
applications, the above approach is usually replaced by a discrete
Monte Carlo simulation of the measurement process, resulting directly
in a system of linear equations for the underlying true
discrete distribution. In this case all the above difficulties are
aggravated by statistical and possibly systematic errors in the
response matrix itself.

In order to avoid these difficulties, it is sometimes advisable to
fold the theoretically predicted true distribution with the estimated
response matrix, and compare the thus folded theoretical spectrum
with the measured one. This method is stable and may be useful
in certain cases, but is clearly useless if a comparison between
various experiments is required, or if the functional form of the
distribution is unknown.

The problem of unfolding has been studied in
various forms, giving rise to a number of independent methods
described in the literature. For instance, a method based on
Bayes' theorem was proposed  in \cite{ALEPH,ARGUS}.
The authors manage to avoid partly the inversion
difficulties by using a non-linear iterative procedure,
leading asymptotically to the unfolded distribution.

Another way of overcoming the instability of unfolding is to use
some kind of regularization condition, based on some {\it{ a priori\/}}
information about the solution. One can demand, for example,
 that the true solution
has minimum curvature (i.e. is quite smooth \cite{Blobel}),
or that it is strictly positive \cite{Schmel}.
These methods usually allow the
suppression of spurious oscillating components of the unfolded
solution and often lead to satisfactory results, though
concrete implementations happen to be quite lengthy and complicated.

In this paper, we propose a new way of analyzing unfolding problems,
based on the {\it{Singular Value Decomposition}} (SVD) of the response
matrix. A simple and transparent 2-dimensional example is used to
explain and illustrate the reasons for the apparent instability of the
problem, and a discrete analog of the minimum curvature condition is
used to stabilize the unfolded solution.
An unfolding procedure based on considerations which are similar to
those presented here has been described
more than a decade ago in \cite{Blobel}, and is still widely used in
experimental analyses. While there are no significant
differences between the foundations of the two procedures,
we believe that our formulation
allows one to obtain more reliable and precise results,
being at the same time much simpler
and easier to implement.

Our approach is based on the extensive use
of SVD, and results in a linear unfolding algorithm which is applicable
to a wide range of problems. We derive also a number of
important recommendations about
the proper normalization of the response  matrix and
choice of the variables. Complete error propagation is implemented,
and a vivid and reliable procedure for determining the optimal value
of the regularization parameter is suggested.

The paper is organized as follows. Sect. {\ref{Notation} contains
our conventions for the notation used throughout the paper. The problem
is formulated in Sect. {\ref{Problem}}. Sect. {\ref{SVD-sec}} is devoted
to the Singular Value Decomposition, illustrated by a simple example of
a $2\times 2$ matrix. The benefits and pitfalls of rescaling
equations and unknowns are analysed in Sect. {\ref{scale}}.
Regularization of the system and actual unfolding is performed in
Sect. {\ref{Unfolding}}, and the choice of the regularization parameter
is discussed in Sect. {\ref{choice}}. The step-by-step unfolding algorithm
is presented in Sect. {\ref{Algorithm}} and is illustrated by two
distinctly different examples in Sect. {\ref{Examples}}. Conclusions are
drawn out in Sect. {\ref{Conclusion}}.

\section{Notation}\label{Notation}

We will adopt the following notation conventions, which will be used
throughout the paper:
\begin{itemize}
\item{All one-dimensional histograms/vectors are denoted by small
letters (e.g. $b,z$ etc.).}
\item{All two-dimensional histograms/matrices are denoted by capital
letters (e.g. $A,X$ etc.).}
\item{The covariance matrix associated with a one-dimensional variable
is denoted by the same letter in capital (i.e. the matrix $W_{ij}$
denotes the covariance matrix of the vector $w_j$).}
\item{No implicit summation is assumed over repeated indices; i.e. no
repeated index is summed unless the summation is explicitly shown.}
\item{Upper index $T$ stands for the transposed matrix, $A^T_{ik}=A_{ki}$,
so that the euclidean norm of a vector $z$ equals $\sqrt{z^Tz}$.}
\item{Upper index $-1$ denotes the inverse matrix, $A^{-1}A=AA^{-1}=I$,
where $I$ stands for the unit matrix, $I_{ik}=\delta_{ik}$.}
\item{All scalar variables are denoted by small greek
letters (e.g. $\e,\tau$ etc.).}
\end{itemize}

\section{The Problem}\label{Problem}

Let the distribution of
a {\it measured} observable be stored in a vector ${ b}$ of dimension $n_b$,
where the $i$th coordinate of the vector contains the number of entries
in the corresponding bin of the histogram.
The measurement is affected by the finite experimental resolution
and/or the limited acceptance of the
detector, so that each event from the true distribution may
find itself
in a range of (not necessarily) adjacent bins, or nowhere at all.
Let us assume, that
we are able to simulate the measurement procedure of this observable
(e.g. with Monte Carlo techniques).
We generate the distribution ${ x}^{\mathrm {ini}}$ of dimension $n_x$,
according to some idea of the underlying physical process,
and perform our detector simulation. At this stage,
every entry in a measured bin (i.e. every {\it event}) can be
directly traced to its origin. This gives us a well
defined system of linear relations between the simulated
{\it true} and {\it measured} distributions:
\begin{equation}\label{xini}
{\hat A}\,{ x}^{\mathrm{ini}}={ b}^{\mathrm {ini} },
\end{equation}

The $n_b\times n_x$ matrix ${\hat A}$ is a probability matrix,
which actually performs the
simulated folding procedure.
Now, with $\hat{A}$ and $x^{\mathrm {ini}}$ given,
for any vector $b$ obtained by a real measurement using the
detector described by its response matrix $\hat{A}$, one can attempt to
find a corresponding unfolded true distribution $x$. It is well known
that trying to solve the linear system of equations
\begin{equation}\label{sys1}
\hat{A}\,x=b
\end{equation}
against $x$ directly, using the exact inversion of the matrix,
 usually leads to completely unacceptable rapidly
oscillating solutions. In the following, based on the {\it
Singular Value Decomposition} (SVD) of the response matrix $\hat{ A}$,
we analyse reasons for this behavior, locate the difficulty and
propose a relatively simple and straightforward regularization method,
which allows one to suppress spurious, quickly oscillating components of
the solution, leaving only statistically significant terms.

The above formulation of the problem may not seem general enough, but it
clearly incorporates most cases of practical interest, is easily
understood and interpreted, and is well-suited for the unfolding method
described in this paper. Let us, however, express the discrete
distributions $x,b$ and the response matrix $\hat A$ in terms of the
underlying continuous probability density functions.

Let $y^{\mathrm {true}}$ be the continuous
true variable under consideration, whose
variation range $\{y^{\mathrm {true}}_0 \div y^{\mathrm {true}}_{n_x}\}$
is divided into $n_x$ bins with boundaries
$y^{\mathrm {true}}_j, j=1,\dots,n_x-1.$
Each component of the vector $x$
is then calculated as an integral over the true distribution function
${\cal{X}}(y^{\mathrm {true}})$ in the appropriate range:
\begin{equation}\label{xj}
x_j=\int_{y_{j-1}^{\mathrm {true}}}^{y_{j}^{\mathrm {true}}}
dy^{\mathrm {true}}
\,{\cal{X}}(y^{\mathrm {true}})\;,
\hspace{1cm}j=1,\dots,n_x.
\end{equation}
Analogously, let $\{y_0 \div y_{n_b}\}$ be the variation range of the
measured variable $y$, with bin boundaries  $y_i, i=1,\dots,n_b-1.$
Then the components of the vector $b$ are appropriate integrals over
the continuous distribution function ${\cal{B}}(y):$
\begin{equation}\label{bi}
b_i=\int_{y_{i-1}}^{y_{i}} dy\,{\cal{B}}(y)\;,\hspace{1cm}i=1,\dots,n_b.
\end{equation}
Let $\hat{\cal{A}}(y,y^{\mathrm {true}})$ be the detector
response function, which maps the
true distribution to the observed one, according to the convolution
integral:
\begin{equation}\label{Axy}
{\cal{B}}(y)=\int_{y^{\mathrm {true}}_{0}}^{y^{\mathrm {true}}_{n_x}}
dy^{\mathrm {true}}\, \hat{\cal{A}}(y,y^{\mathrm {true}}) \,
{\cal{X}}(y^{\mathrm {true}})\;.
\end{equation}
After this, the response matrix $\hat{A}$ can be defined as the ratio
of two integrals:
\begin{equation}\label{Arat}
\hat{A}_{ij}=\frac{{\int_{y_{i-1}}^{y_{i}} dy
{\int_{y^{\mathrm {true}}_{j-1}}^{y^{\mathrm {true}}_{j}}
dy^{\mathrm {true}} \,\hat{\cal{A}}(y,y^{\mathrm {true}})
\,{\cal{X}}(y^{\mathrm {true}})}} }
{{\int_{y^{\mathrm {true}}_{j-1}}^{y^{\mathrm {true}}_{j}} }
dy^{\mathrm {true}} \,{\cal{X}}(y^{\mathrm {true}})}\;.
\end{equation}
Each element $\hat{A}_{ij}$ equals to the probability for an event
generated in the {\it true} bin $j$ to be found in {\it measured} bin $i$.

In high energy physics applications the response function
$\hat{\cal{A}}(y,y^{\mathrm {true}})$ is
usually not known analytically.
Instead, some sophisticated
detector simulation techniques are used to determine the matrix directly
as explained above.

\section{Singular Value Decomposition}\label{SVD-sec}

\subsection{Definitions}

A singular value decomposition (SVD) of a real $m\times n$ matrix $A$ is
its factorization of the form
\begin{equation}\label{svd}
A = U\,S\,V^T
\end{equation}
where $U$ is an $m\times m$ orthogonal matrix,
$V$ is an $n\times n$ orthogonal matrix, while S is an $m\times n$
 diagonal matrix with non-negative diagonal elements:
\begin{equation}\label{UV}
U\,U^T=U^TU=I, \;\;\;V\,V^T=V^TV=I,
\end{equation}
\begin{equation}\label{sij}
S_{ij}=0\ \ {\mathrm{for}}\ \ i \neq j, \;\;\; S_{ii} \equiv s_i \geq 0.
\end{equation}
The quantities $s_i$ are called {\it singular values} of the matrix A, and
columns of $U$ and $V$ are called the left and right {\it singular vectors}.

The singular values contain very valuable information about the properties
of the matrix. If, for example, $A$ is itself orthogonal, all its singular
values are equal to 1. On the contrary, a degenerate matrix will have
at least one zero among its singular values. In fact, the rank of a
matrix is the number of its non-zero singular values. If the matrix and/or
the r.h.s. of a linear system is known with some level of uncertainty,
and some singular values of the matrix are significantly smaller than
others, the system may be difficult to solve even if formally the
matrix has full rank. In many aspects such matrices behave like
degenerate ones, and SVD suggests a method of treating such problems,
which is common for small and exactly zero singular values.

We will assume that the singular values $s_i$ form a
non-increasing sequence. This is easily achieved by swapping pairs of
singular values, swapping simultaneously corresponding columns of $U$ and $V$.
We will assume also that $m \geq n$, which means that the number of bins in
the measured histogram $b$ should not be smaller than the number of bins in
the unfolded histogram $x$. If necessary, one can just add rows of zeroes to
the initial matrix.

Comprehensive descriptions of SVD with many technical
details and examples can
be found in the literature (see, e.g., the excellent books \cite{SVD1,SVD2}).
One of the most attractive features of this procedure is that one
does not really have to perform SVD by hand. A very efficient and transparent
FORTRAN subroutine (called, not surprisingly, {\it SVD}) is present in
the CERN program library. Some earlier implementations can be found in
refs. \cite{SVD1,SVD2} as well.

Once the matrix
is decomposed into the form (\ref{svd}),
its properties can be readily analyzed
and it becomes very easy to manipulate, as illustrated in following
subsections. This kind of analysis is extremely useful for ill-defined
linear systems with almost (or even exactly) degenerate matrices,
as it not only locates the difficulty, but can also suggest
ways of overcoming it.

\subsection{A simple example}\label{Ex_SVD}

Consider a very simple $2\times 2$ example, which
nevertheless incorporates most of the interesting aspects of the problem.

Let the response matrix $\hat{A}$ have the form
\begin{equation}\label{a22}
\hat{A} = {1\over 2} \left(
\begin{array}{cc}
1 + \e & 1 - \e \\
1 - \e & 1 + \e
\end{array}
\right),
\end{equation}
with $0 \leq \e \leq 1$ determining the "quality" of the detector:
$\e=1$ means an ideal detector with the response matrix equal to
unity, while small $\epsilon \ll 1$ corresponds to a poor detector,
almost unable to distinguish the two bins. Note however, that the
overall efficiency is 100\%, so that no event escapes detection
(sum of elements in each column equals 1). The measurement process
now is simulated by multiplying the matrix $\hat{A}$
over the true
distribution $x$, resulting in the measured histogram $b$:
\begin{equation}\label{Axb}
\hat{A}\,x = b
\end{equation}
With  vector $b$ measured and the response matrix (\ref{a22}) given,
one can try to unfold the true distribution.

Singular value decomposition of $2\times2$ matrices is very simple,
involving just a single rotation from left and another from right.
As the matrix (\ref{a22}) is symmetric, the orthogonal matrices $U$ and
$V$ should coincide. SVD can be performed explicitly by hand, and
one easily obtains:
\begin{equation}\label{svd22}
\hat{A} = U\,S\,V^T
\end{equation}
with
\begin{equation}\label{usv22}
U=V={1\over {\sqrt2}} \left(
\begin{array}{cc}
1  & 1  \\
1  & -1
\end{array}
\right),\;\;\;\;\;\;\;\;
S=\left(
\begin{array}{cc}
1  & 0  \\
0  & \e
\end{array}
\right).
\end{equation}
So, the two singular values of the matrix (\ref{a22}) are $s_1=1,\ s_2=\e$.

\subsection{Solving a linear system using SVD}\label{sol}

Suppose now that the apparatus described by the matrix (\ref{a22}) has
been used to measure numbers of events in a two-bin histogram
\begin{equation}\label{b}
b = \left(
\begin{array}{c}
  b_1 \\
  b_2
\end{array}
\right).
\end{equation}
Let $B$ be the corresponding covariance matrix,
which is especially simple for purely statistical errors
in independent entries $b_1$ and
$b_2$:
\begin{equation}\label{B2}
B = \left( \begin{array}{cc} b_1 & 0 \\ 0 & b_2 \end{array}\right).
\end{equation}

In order to solve the system, let us use $U,S$ and $V$ from
(\ref{usv22}) to rotate
both the unknown vector $x$ and the r.h.s. of the system $b$,
\begin{equation}\label{zd}
z=V^T\,x = {1\over{\sqrt{2}}} \left(
\begin{array}{r}
  {x_1+x_2}\\
  {x_1-x_2}\\
\end{array}
\right),\;\;\;\;\;\;\;\;\;\;
d=U^T\,b = {1\over{\sqrt{2}}} \left(
\begin{array}{r}
  {b_1+b_2}\\
  {b_1-b_2}\\
\end{array}
\right),
\end{equation}
in order to form a {\it diagonal} system of equations
\begin{equation}{\label{Szd}}
S\,z=d,\;\;\;\;\;\;\; z=S^{-1}d,
\end{equation}
where
\begin{equation}\label{s-1}
S^{-1}=\left(
\begin{array}{cc}
1  & 0  \\
0  & {1\over{\e}}
\end{array}
\right).
\end{equation}
The unknown vector $x$ can now be easily obtained by rotating $z$
back:
\begin{equation}{\label{xVz}}
x=Vz=VS^{-1}d=VS^{-1}U^Tb=\hat{A}^{-1}b=
{{b_1-b_2}\over{2\e}}\left(
\begin{array}{c} {1 }\\{-1}\end{array}
\right)
+{{b_1+b_2}\over{2}}\left(
\begin{array}{c} {1 }\\{1}\end{array}
\right).
\end{equation}
Expression (\ref{xVz}) gives the exact solution of the system (\ref{Axb})
for whatever small but finite $\e$.

Formally, SVD of the matrix $\hat A$ means a decomposition
of the r.h.s. $b$ into a series of
orthogonal and normalized functions of the discrete variable
$i=1,\dots,n_b$. The basis is given by the
columns of the matrix $U$, and the components of the vector $d$
form the coefficients of this decomposition. Similarly, the vector
of unknowns $x$ is also decomposed into a series of ortho-normalized
functions of the discrete variable $j=1,\dots,n_x$, given by the
columns of the matrix $V$, while the coefficients stored in
the vector $z$ are new unknowns. After performing
these transformations, the initial
system of equations (\ref{Axb}) is reduced to the diagonal system (\ref{Szd})
which can be easily solved: the matrix $S$ in (\ref{usv22}) is diagonal
and can be inverted by just inverting the singular values.

The inverse matrix $\hat{A}^{-1}$
exists for any $\e\ne 0$:
\begin{equation}\label{A-1}
\hat{A}^{-1} = V\,S^{-1}U^T= {1\over {2\e}} \left(
\begin{array}{rr}
  1 + \e & - 1 + \e \\
- 1 + \e &   1 + \e
\end{array}
\right)\equiv
{1\over 2}\left(
\begin{array}{cc}  1  &  1\\  1  &  1  \end{array}
\right)
 + {1\over {2\e}} \left(
\begin{array}{cc}  1  & - 1\\- 1  &   1  \end{array}
\right).
\end{equation}
Note that the expressions (\ref{xVz}) and
(\ref{A-1}) are exact, so that SVD and the
subsequent analysis can be considered as just another method of solving
 well determined full rank linear systems, maybe a bit too complicated
but quite capable.
If all components of the rotated r.h.s. $d$ are statistically
significant {\it and} if neither of the singular values $s_i$
of the matrix $\hat{A}$ is too small,
the system (\ref{Axb}) can be solved without any
problem using any other method like, say, Gaussian elimination.
But if $\e$ is small
the problem becomes ill-determined,
and when in addition the r.h.s. is affected by measurement errors,
the exact solution usually does not make any sense.
In this case clearly it is not the exact solution we are looking for,
and conventional methods of solving linear systems do not work.
Usually they cannot even detect the problem.

To illustrate this, let us assume that the measured event numbers
$b_1$ and $b_2$ satisfy the following relation:
\begin{equation}\label{acc}
(b_1-b_2)^2 \leq b_1+b_2.
\end{equation}
This means that the difference $b_1-b_2$ is not
statistically significant, so that the first term in the exact
solution (\ref{xVz})
is in fact a random number. But if $\e$ is small enough
(in this case - smaller than $1/\sqrt{b_1+b_2}$), this first term
in both $x_1$ and $x_2$ dominates over the well-behaved and statistically
significant second term, leading to almost arbitrary and senseless results.
This phenomenon can be easily understood: for very small $\e$ the apparatus
is almost "blind", and one can hardly expect to determine $x_1$ and $x_2$
separately, unless the errors in $b$ are sufficiently small.

\subsection{Locating the difficulty}\label{Locate}

In order to trace the problem back to its origin let us reconsider the
rotated system (\ref{Szd}).
Under the assumption (\ref{acc}) about the statistical accuracy of the data,
the covariance matrix $D$ of the rotated r.h.s. $d$ is approximately
diagonal,
\begin{equation}\label{D}
D =U^TB\,U = {1\over 2} \left(
\begin{array}{cc}
  b_1+b_2 & b_1-b_2 \\
  b_1-b_2 & b_1+b_2
\end{array}
\right)\approx {1\over 2} \left(
\begin{array}{cc}
  b_1+b_2 & 0 \\
        0 & b_1+b_2
\end{array}
\right),
\end{equation}
so one has the following (almost) independent set of equations:
\begin{eqnarray}\label{z1}
1\cdot z_1&           &=(b_1+b_2)/\sqrt{2}\pm\sqrt{(b_1+b_2)/2},\\
\label{z2}
          &\e\cdot z_2&=(b_1-b_2)/\sqrt{2}\pm\sqrt{(b_1+b_2)/2}.
\end{eqnarray}
The first equation is good and gives a sensible result for $z_1$, but
the second one is not too useful even if $\e$ is
large, as the r.h.s. is in fact a random number. However, for $\e$
close to unity
it is at least harmless; it becomes dangerous for small $\e$, when the
random number in the r.h.s. gets strongly amplified after being divided
by $\e$, and after the
rotation back to $x$ according to (\ref{xVz}),
gives a huge and senseless contribution to {\it both}
components of $x$ in the exact solution (\ref{xVz}).
This means that $z_2$ cannot and should not be determined from
equation (\ref{z2}),
because effectively the matrix has insufficient rank and the system is
over-determined. However sensible equation (\ref{z2}) may look,
it should be replaced by another equation
\begin{equation}\label{zero}
0\cdot z_2 = 0\pm\sqrt{(b_1+b_2)/2},
\end{equation}
which contains as much information as (\ref{z2}), but is much less
harmful. The value of $z_2$ is completely arbitrary and can be determined
only from some external condition. Note that each value
of $z_2$ will lead to a new vector $x$.
The easiest thing to do is to put $z_2=0$,
which would lead to the "shortest" $x$:
$\parallel x\parallel^2=x_1^2+x_2^2\equiv z_1^2+z_2^2$, as orthogonal
transformations do not change the euclidean norm of a vector.
Or, alternatively, one could choose the solution minimizing the
variation $(x_1-x_2)^2$. In our small example both these
alternatives lead to the same "regularized" vector
\begin{equation}\label{xreg}
x^{\mathrm {reg}}={{b_1+b_2}\over{2}}\left(
\begin{array}{c} {1 }\\{1}\end{array}
\right).
\end{equation}
Corresponding regularized covariance matrix has the form
\begin{equation}\label{Xreg}
X^{\mathrm {reg}}={{b_1+b_2}\over{4}} \left(
\begin{array}{rr} 1&1\\1&1\end{array}
\right).
\end{equation}
These two equations define the best reasonable estimate for an unfolded
vector $x$, given the condition on the statistical accuracy of the
data  (\ref{acc}) and a poor detector with $\e\ll 1/\sqrt{b_1+b_2}$.

\section{Rescaling equations and normalizing unknowns}\label{scale}

Let us now get back to the full-scale problem defined in
Sect. \ref{Problem} and look at the
initial linear system (\ref{sys1}) from
another viewpoint. It represents the solution of the following
least square problem:
\begin{equation}\label{chi1}
\sum_{i=1}^{n_b}{\bigl(
{\sum_{j=1}^{n_x} \hat{A}_{ij}x_j-b_i}} \bigr)^2 = {\mathrm {min}},
\end{equation}
and is adequate if the equations are exact, or if the errors in the r.h.s.
are identical.
This is not generally the case, as
measurement errors
in the vector $b$ vary from bin to bin, and hence, different
equations have different significance. In fact, one should consider
a weighted least squares problem, where the following
expression is being minimized:
\begin{equation}\label{chi2}
\sum_{i=1}^{n_b}{\bigl(
{{\sum_{j=1}^{n_x} \hat{A}_{ij}x_j-b_i}\over{\Delta b_i}}} \bigr)^2
= {\mathrm {min}},
\end{equation}
where $\Delta b_i$ is the error in $b_i$.
The general case of (\ref{chi2}) reads
\begin{equation}\label{chi3}
(\hat{A}x-b)^T B^{-1} (\hat{A}x-b) = {\mathrm {min}}~,
\end{equation}
where $B$ is the covariance matrix of the measured vector $b$.

\subsection{Normalization of the unknowns}\label{scalx}

It is well known, that
the exact solution of a well-determined linear system remains
unchanged, if
either the equations, or the unknowns, or both are
rescaled. However, in the cases under consideration
(where $n_x \le n_b$ and some singular values are small)
the minimization of (\ref{chi3}) leads to an
overdetermined system which should be solved in the least-squares sense.
In this case any rescaling of equations and/or unknowns changes the
singular values of the system and hence the solution as well.
One can suggest various ways of rescaling, and some of them may lead
to a serious improvement in the system behavior. One of our
most important tasks is to optimize the system by rescaling it so
that significant
information is not suppressed while non-significant is not enhanced.

One can, for instance, divide an unknown $x_k$ everywhere in the system
by a number $\lambda$, multiplying simultaneously all corresponding
coefficients $A_{ik}, i=1,\dots,n_b$ by the same number $\lambda$.
Choosing various $\lambda$'s for different $k$'s, one can obtain
substantially different matrices.
We argue that one particular
choice of rescaling coefficients is the most suitable for our
purposes, provided the probability matrix $\hat A$ is obtained using
a Monte Carlo simulation procedure (see Sect. \ref{Problem}).

Consider a new unknown vector $w_j=x_j/x^{\mathrm{ini}}_j$, which
measures the {\it {deviation}} of $x$ from the initial Monte Carlo input
vector $x^{\mathrm{ini}}$. If one now multiplies each column of the
probability matrix $\hat A_{ij}$ by the corresponding number of events
generated in this bin $x^{\mathrm{ini}}_j$, the system becomes
\begin{equation}\label{sys2}
\sum _{j=1}^{n_x}A_{ij}w_j=b_i,
\end{equation}
where $A_{ij}$ is no longer the {\it{probability,}} but rather the {\it
actual number of events,} which were generated in bin $j$ and ended up
in bin $i$
{\footnote{
If defined through continuous probability distributions, this new matrix
is equal to the {\it {numerator}} of eq. (\ref{Arat}).}}.
Obviously, $x=x^{\mathrm{ini}}$ corresponds to all components of the
vector $w$ being equal to 1, so that
$b_i^{\mathrm{ini}}=\sum _{j=1}^{n_x}A_{ij}.$
At the end of the unfolding procedure,
in order to obtain the correctly normalized
unfolded solution $x_j$, one has to multiply the unfolded vector
$w$ by $x^{\mathrm{ini}}$:
\begin{equation}\label{xwx}
x_j=w_jx^{\mathrm{ini}}_j,\;\;\;\;\;\;\;\;j=1,\dots,n_x.
\end{equation}
Of course, if the number of generated events is the same for each bin,
$x^{\mathrm{ini}}=\mathrm{const}$, then the probability matrix $\hat A$
and the number-of-events matrix $A$ coincide up to an overall constant
factor which is completely irrelevant for our analysis.

The systems (\ref{sys1}) and (\ref{sys2}) are completely
equivalent for any shape of $x^{\mathrm{ini}}$, if the exact solution
is required, but there are two serious reasons why, for the class of
problems considered here, (\ref{sys2}) is much better suited.

The first reason is fairly simple:
if the initial Monte Carlo distribution $x^{\mathrm{ini}}$ is
physically motivated and is reasonably close to the one being unfolded,
the unknown vector $w$
should be smooth and should have small bin-to-bin variation,
thus requiring less terms in the
decomposition into orthogonal functions. This in turn means that more
accurate unfolding should be possible, as fewer unknowns are required
in order to obtain the unfolded solution.

The second reason is more technical and is connected to the singular
value analysis. Whatever high statistics is generated in order to
obtain the matrix $ A$, some of its columns and/or rows may contain
very few events, and some elements
$ A_{ij}$ may have just a single entry.
In the probability matrix, these elements will contain {\it{the
largest possible value}} of 1, unjustifiably giving a high weight to that
particular equation and unknown, and the fact that this
element has a 100 \% error is completely ignored. At the same time,
highly populated columns with statistically well-determined elements
usually contain values significantly smaller than 1, due to finite
resolution and limited acceptance. This clearly makes the
probability matrix $\hat A$
a bad choice. On the contrary, the elements of the
number-of-event matrix $A$ are large if the generated statistics
is large, and vice versa, thus giving a larger weight to better
determined equations and unknowns.

The latter argument is in fact
based on a quite formal consideration
using perturbation theorems for the singular values of a matrix \cite{SVD2},
and is an attempt to account for the intrinsic errors in the response
matrix. It can be shown that for the cases of interest, initial Monte
Carlo statistics should be at least one or two orders of magnitude higher
than the statistics of the measured data. If so, the error in the
unfolded solution based on (\ref{sys2}) is dominated by the
measurement errors in the
r.h.s. $b$. Note that this is not true for the system based on the
probability matrix (\ref{sys1}), where huge errors may be introduced
because of the fact that scarcely populated areas of the response matrix
have far larger weight than they deserve.

\subsection{Rescaling equations}\label{scalr}

The very form of (\ref{chi2}) clearly suggests the way
of rescaling the equations:
after dividing each equation by
the corresponding error $\Delta b_i$ one obtains a balanced system,
where all the equations have equal weights.

If $B$ is not diagonal, equation rescaling
becomes slightly more complicated but still straightforward.
Being a covariance matrix,
$B$ should be symmetric and positive-definite, so its SVD yields:
\begin{equation}\label{B}
B=QRQ^T,\;\;\;\;\;R_{ii}\equiv r_i^2 \neq 0,\;\;\;\;\;
R_{ij}=0~{\mathrm{for}}~i\neq j,\;\;\;\;\;B^{-1}=QR^{-1}Q^T.
\end{equation}
Substituting $B^{-1}$ into (\ref{chi3}) one sees that after the
rotation and rescaling of both the r.h.s. $b$ and the matrix $A$,
\begin{equation}\label{rota}
\tilde{A}_{ij}={1\over{r_i}} {\sum_{m}Q_{im}A_{mj}},\;\;\;
\tilde b_{i}={1\over{r_i}} {\sum_{m}Q_{im}b_{m}},
\end{equation}
the expression being minimized looks very simple again,
\begin{equation}\label{chi9}
(\tilde Aw-\tilde b)^T(\tilde Aw-\tilde b)={\mathrm{min}},
\end{equation}
and the minimization leads to the following system:
\begin{equation}\label{nsa}
\sum_j\tilde A_{ij}w_j=\tilde b_i.
\end{equation}
The covariance matrix of the rescaled r.h.s., $\tilde B$, is now explicitly
made equal to the unit matrix $I$, and all the equations have equal
importance.

\section{Regularization and unfolding}\label{Unfolding}

The transition from (\ref{chi1}) to (\ref{chi9}) changes the appearance
of the system from (\ref{sys1}) to (\ref{nsa}). The singular values
of the matrix are also changed, but the main
problem with small singular values still remains.
The exact solution of (\ref{nsa}) will again most certainly lead to a rapidly
oscillating distribution, which may have a smaller amplitude but is still
 useless. This spurious oscillatory
component should be suppressed, using some {\it {a priori}} knowledge
about the solution. Technically this can be achieved by adding the
{\it{regularization}} or {\it{stabilization}}
term to the expression to be minimized (see \cite{SVD2,Blobel,Schmel}
and references therein):
\begin{equation}\label{chi5}
(\tilde A\,w-\tilde b)^T\, (\tilde A\,w-\tilde b) +
\tau\cdot(C\,w)^T C\,w\;=\; {\mathrm {min}}.
\end{equation}
Here $C$ is a matrix which defines the
{\it {a priori}} condition on the solution,
while the value of the regularization
parameter $\tau$ determines the relative weight of
this condition. For example, the choice $C_{ik}=\delta_{ik}$ would
minimize the euclidean norm of the vector $w$, and if $\tau$ is set to be
infinitely large, this would result in a vector $w_j=0$ for any $\tilde A$
and $\tilde b$.

While the optimal value of $\tau$ is very much problem-dependent and its
determination is an important part of our procedure, the explicit form
of the matrix $C$ should be chosen from general considerations. The common
belief is that the solution histogram $w$ should be smooth, with small
bin-to-bin variation. Let us define the "curvature" of the discrete
distribution $w_j$ as the sum of the squares of its second derivatives:
\begin{equation}
\sum_{i} [(w_{i+1}-w_i)-(w_i-w_{i-1})]^2\;.
\end{equation}
Then the choice
\begin{equation}\label{cik}
C=\left(
\begin{array}{cccccc}
-1 & 1 & 0 & 0 & \dots &   \\
 1 &-2 & 1 & 0 & \dots &   \\
 0 & 1 &-2 & 1 & \dots &   \\
   &\dots& &   & \dots &   \\
   &\dots& & 1 & -2    & 1 \\
   &\dots& &   &  1    &-1  \end{array} \right)
\end{equation}
will suppress solutions $w$ having large curvatures. Minimization of
(\ref{chi5}) leads to a new linear system, which has $n_x$ additional
equations:
\begin{equation}\label{regs}
\left[
\begin{array}{c} {\tilde A}\\{{\sqrt{\tau}}\cdot C}\end{array}
\right]w
=\left[
\begin{array}{c} {\tilde b }\\{0}\end{array}
\right].
\end{equation}
This system is clearly over-determined, and one can apply SVD to the
$(n_b+n_x)\times n_x$ matrix in the l.h.s. in order to solve it. This
is possible, but would require calling SVD for each value of $\tau$.
Fortunately, a more efficient method (called sometimes {\it {damped
least squares}} \cite{SVD2} )  can be suggested, which allows to express
the solution of (\ref{regs}) for any $\tau$ through the solution of
the initial non-regularized problem corresponding to $\tau=0$.
The first step is to make the regularization term
proportional to the unit matrix $I$:
\begin{equation}\label{rots}
\left[
\begin{array}{c} {\tilde A\,C^{-1}}\\{{\sqrt{\tau}}\cdot I}\end{array}
\right]C\,w
=\left[
\begin{array}{c} {\tilde b }\\{0}\end{array}
\right].
\end{equation}
For $\tau=0$ the system (\ref{rots}) is equivalent to (\ref{nsa}), if
the inverse $C^{-1}$ exists and can be safely calculated. The "second
derivative" matrix (\ref{cik}), however, is apparently degenerate
(every column and every row sums up to zero) so some measures should be
taken to make the exact inversion possible. The easiest thing to do is
to add a small diagonal component,
$C_{ik}\Rightarrow C_{ik}+\xi\cdot\delta_{ik}$, with $\xi$ large enough
to make the inversion possible, but small enough not to change
significantly the condition of minimum curvature. In most cases,
$\xi=10^{-3}$ or $10^{-4}$ is a good choice. $C$ now is a symmetric
non-singular matrix,
\begin{equation}\label{cik1}
C=\left(
\begin{array}{cccccc}
-1+\xi & 1 & 0 & 0 & \dots &   \\
 1 &-2+\xi & 1 & 0 & \dots &   \\
 0 & 1 &-2+\xi & 1 & \dots &   \\
   &\dots& &   & \dots &   \\
   &\dots& & 1 & -2+\xi    & 1 \\
   &\dots& &   &  1    &-1+\xi  \end{array} \right)
\end{equation}
which can be inverted using standard techniques
{\footnote{One has, however, to be careful, as the matrix is too close
to a singular one, and some of the standard routines may not work for
small $\xi$. E.g., RFIN, RSINV and RSFINV have failed for $\xi=10^{-4}$,
while RINV was successful.
So, it may be convenient to use SVD once again for
this purpose:  decompose $C=U_CS_CV^T_C$ and then calculate
$C^{-1}=V_CS^{-1}_CU^T_C$.}}.

Let us now solve the system (\ref{rots}) with $\tau=0$.
First, one needs SVD to decompose the product of matrices
$\tilde A\,C^{-1}$:
\begin{equation}\label{ac-1}
\tilde A\,C^{-1}= U\,S\,V^T.
\end{equation}
Here, once again, $U$ and $V$ are orthogonal and $S$ is diagonal, with
non-increasing positive diagonal elements $s_i$.
The solution now proceeds as in the $2\times2$ example of section
\ref{Ex_SVD}. Rotate both $\tilde b$ and $Cw$ to obtain a diagonal system:
\begin{equation}\label{d}
d\equiv U^T\tilde b\;,\;\;\;\;\;\;z\equiv V^TC\,w.
\end{equation}
The system now looks (and actually is) very simple:
\begin{equation}\label{szd}
s_i\cdot z_i = d_i\;,\;\;\;\;\;\;i=1,\dots,n_x.
\end{equation}
Note that because the covariance matrix of the r.h.s. $\tilde b$ was made
equal to the unit matrix, the orthogonality of $U$ guarantees that the
new rotated r.h.s. $d$ also has a unit covariance matrix, i.e. the
equations in (\ref{szd}) are completely independent and have identical
unit errors in their r.h.s.

Obviously, solving (\ref{szd}) one obtains the exact solution of the
non-regularized system:
\begin{equation}\label{zwo}
z_i^{(0)} = {{d_i}\over{s_i}},\;\;\;\;\;\;w^{(0)}=C^{-1}Vz^{(0)}
\end{equation}
and the true distribution $x$ can be obtained by multiplying each $w_i$ by
the corresponding $x_i^{\mathrm{ini}}$. With $\tau=0$ there is no
regularization, so this solution is as useless as it used to be. But the
solution of the system (\ref{regs}) with nonzero $\tau$ can now be found
very easily, using the procedure explained in detail in
\cite{SVD2} (Chapter 25, Sect.4).
In short, introducing non-zero $\tau$ is effectively equivalent to changing
 $d_i$ by a regularized distribution:
\begin{equation}\label{dt}
d_i^{(\tau)} = d_i{{s_i^2}\over{s_i^2+\tau}},
\end{equation}
so that the solution of the rotated system becomes
\begin{equation}\label{zwt}
z_i^{(\tau)} = {{d_i\,s_i}\over{s_i^2+\tau}},\;\;\;\;\;\;
w^{(\tau)}=C^{-1}V\,z^{(\tau)}.
\end{equation}
One can now see how nonzero $\tau$  regularizes the singularities
due to small $s_i$'s, effectively working as a cutoff for a low-pass
filter, if Fourier-transform terminology is used. Indeed, $s_i$
is small when  the index $i$ is large, which in general
corresponds to quickly oscillating singular vectors (i.e. columns
of $U$ and $V$) defining the new basis in the rotated space.

Continuing the analogy with Fourier analysis, one can mention that the
cutoff provided by the above regularization procedure happens to be
quite smooth, thus avoiding specific quasi-periodic fluctuations of
the solution known as the Gibbs phenomenon.

The covariance matrices $Z$ and $W$ of the solutions (\ref{zwt})
can now be easily calculated:
\begin{eqnarray}\label{ZWt}
Z_{ik}^{(\tau)}&=&{{s_i^2}\over{(s_i^2+\tau)^2}}\cdot\delta_{ik}, \\
W^{(\tau)}&=&C^{-1}V\,Z^{(\tau)}V^T\,C^{T-1}.
\end{eqnarray}

Now in order to obtain the true unfolded distribution $x$ and its covariance
matrix $X$ one has to multiply $w$ and $W$ by the initial Monte Carlo
distribution $x^{\mathrm{ini}}$:
\begin{eqnarray}\label{xX}
x_{i}^{(\tau)}&=&x_i^{\mathrm{ini}}\,w_{i}^{(\tau)}, \\
\label{Xx}
X_{ik}^{(\tau)}&=&x_i^{\mathrm{ini}}\,W_{ik}^{(\tau)}x_k^{\mathrm{ini}}.
\end{eqnarray}

It is important to note that while (\ref{xX}) and (\ref{Xx}) are regularized
and as such depend on the value of $\tau$, the {\it{inverse}} of the
covariance matrix $X^{-1}$ (which should be used for any $\chi^2$
calculation involving the unfolded distribution (\ref{xX})),
is regular and readily calculable:
\begin{equation}\label{Xinv}
X^{-1}_{jk}={1\over{x^{\mathrm {ini}}_jx^{\mathrm {ini}}_k}}\,
{\sum_i\tilde A_{ij}\tilde A_{ik}}.
\end{equation}
In fact, $X^{(\tau)}$ defined by (\ref{Xx}) is the {\it effective
pseudoinverse} of the matrix (\ref{Xinv}). This means that while
the equation
\begin{equation}\label{XX1}
X^{(\tau)}X^{-1}X^{(\tau)}=X^{(\tau)}
\end{equation}
is valid as if $X^{(\tau)}$ were the true inverse of $X^{-1}$,
for a different ordering one has (see \cite{SVD1,SVD2}):
\begin{equation}\label{XX2}
\parallel X^{-1}X^{(\tau)}X^{-1}-X^{-1}\parallel < \tau.
\end{equation}

It may be interesting to write out the exact inverse covariance matrix
for the $2\times 2$ example of Sect. {\ref{SVD-sec}}:
\begin{equation}\label{X-1}
X^{-1}=\hat{A}^TB^{-1}\hat{A}=
{1 \over 4}\left({1\over b_1}+{1\over b_2}\right) \left(
\begin{array}{rr}
1+\e^2  & 1-\e^2  \\
1-\e^2  & 1+\e^2
\end{array}
\right)+
{\e \over 2}\left({1\over b_1}-{1\over b_2}\right) \left(
\begin{array}{rr}
1  & 0  \\
0  & -1
\end{array}
\right).
\end{equation}
The regularized covariance matrix (\ref{xreg}) is actually the pseudoinverse
of (\ref{X-1}). As expected, the latter is perfectly regular in the limit
of zero $\e$.

\section{Error analysis and choice of $\tau$}\label{choice}

Very important and interesting information about the whole problem can
be disclosed by plotting $d_i$, or, better, ${\mathrm{log}}|d_i|$ vs $i$.
As mentioned in subsect. \ref{sol},
the $i$-th component of the vector $d$ is
the coefficient in the decomposition of the measured (and rescaled)
histogram $\tilde b$ in front of a basis function defined by the
$i$-th column of the rotation matrix $U$. For reasonably smooth
measured distributions, only the first few (say, $k$)
 terms of the decomposition are
expected to be significant, while the contribution of quickly
oscillating basis vectors corresponding to large values of $i>k$ should
be compatible with zero, well within the
statistical errors in $d_i$ (which are equal to 1 for
all $i$).
So, on the plot one should see two separate patterns (see section
\ref{Examples} for a few illustrations):
for small $i$, $d_i$ should be statistically significant,
$|d_i|\gg 1$, falling gradually (usually exponentially)
towards a gaussian-distributed random value for large $i$ with the
variance equal to 1 and the mean close to zero
(the absolute values of non-significant
components  $|d_i|, i>k$\/ should have the average close to
$1/2{\sqrt{\pi}}\approx 0.28$).
The critical value $i=k$, after which $d_i$'s
are non-significant, determines the effective rank of the obtained
system of equations.
Usually it is clearly seen on the plot of
${\mathrm{log}}|d_i|$ vs $i$, as the value of $i$ where the behavior of
$d_i$ changes from exponentially falling to a constant.

The standard statistical tests can be used to check whether
the last $n_x-k$\/ components of $d_i$ are compatible with the expected
normal distribution with zero mean and unit variance. If this is not the case,
then the errors in the measured data (or maybe the
response matrix itself) are not estimated correctly, and we
recommend finding out the reason of the discrepancy before proceeding with
the unfolding.
If, for example,the actual measurement errors in $b$ are
under(over)estimated,  then
the variance of $d_i$ for $i>k$ will be smaller (larger) than 1.
Moreover, if some additional correlations exist in the measured data
 which are not accounted
for in the covariance matrix $B$, then ${\mathrm{log}}|d_i|$ may steadily
decrease for all $i$, though probably for $i>k$ the slope will be different.

All this shows that the analysis of the plot of
${\mathrm{log}}|d_i|$ vs $i$ is of great interest by itself, being able
to reveal the actual level of understanding the measurement errors in the
experiment described by the simulated matrix $A$.

Anyway, if the number of statistically significant equations is
determined to be equal to $k$, the regularization parameter $\tau$
should be put equal to the square of the $k$th singular value
$s_k\equiv S_{kk}$ of the
matrix $\tilde AC^{-1}$, determined in (\ref{ac-1}):
\begin{equation}\label{tau}
\tau=s_k^2.
\end{equation}
With $\tau$ given by (\ref{tau}), the unfolded vector $x$, its covariance
matrix $X$ and the inverse of the latter $X^{-1}$ are completely defined
by corresponding equations (\ref{xX}),(\ref{Xx}) and (\ref{Xinv})
and can be easily calculated, forming the solution of the unfolding
problem.

Yet another (and maybe more convincing) way of determining $\tau$ is to
generate a test distribution which is close to the expected true one, but
still significantly different from the initial Monte Carlo distribution
$x^{\mathrm{ini}}$. Then one should
simulate the measurement process by applying the response matrix to it,
and add corresponding random statistical errors to the thus
obtained "measured" distribution. The described unfolding procedure
should be applied to the latter, and the best choice for $\tau$ is the one
giving the smallest $\chi^2$ between the test and the unfolded
distributions (see our second example in Sect. {\ref{Examples}}).

\section{The algorithm}\label{Algorithm}

In this section we present the concise description of the complete
unfolding algorithm. The algorithm is linear (i.e. contains no loops)
and can be divided into three distinct parts: initialization,
rescaling/rotation and actual unfolding. Each step includes references
to relevant subsections and equations.

\begin{itemize}
\item{Initialization:}{

\begin{enumerate}
\item {Define the number of bins $n_b$ and bin boundaries of the measured
histogram $b$.}
\item {Define the number of bins $n_x$ and bin boundaries, common for the
initial Monte Carlo $x^{\mathrm {ini}}$ and the unfolded distribution $x$.}
\item {Build the "second derivative" matrix $C$,
according to eq. (\ref{cik1}).}
\item {Calculate the inverse $C^{-1}$ (see Sect.{\ref{Unfolding}}).}
\item {Generate the initial Monte Carlo histogram $x^{\mathrm {ini}}$, and
simulate the detector response in terms of the
two-dimensional $n_b\times n_x$ histogram $A$.
Elements of $A$ should contain actual
numbers of events, rather than probabilities.}
\item {Read and fill the measured distribution $b$ and its covariance
matrix $B$.}
\end{enumerate}
}
\item{Rescaling and rotation:}{

\begin{enumerate}
\item {Perform SVD of the covariance matrix $B$, according to eq. (\ref{B}).}
\item {Rotate and rescale both the r.h.s. $b$ and the matrix $A$, in order
to make the covariance matrix of the r.h.s. equal to the unit
matrix, according to eqs. (\ref{rota}).}
\item {Calculate the inverse of the covariance matrix, $X^{-1}$,
 of the unfolded vector $x$, according to eq. (\ref{Xinv})}.
\item {Multiply matrices $\tilde A$ and $C^{-1}$ and perform SVD of the
product, according to eq. (\ref{ac-1}).}
\item {Calculate the rotated r.h.s. $d$, according
to eq. (\ref{d}).}
\end{enumerate}
}
\item{Unfolding:}{

\begin{enumerate}
\item {Plot ${\mathrm{log}}|d_i|$ vs $i$
and determine the effective rank $k$ of the system (see Sect. \ref{choice}).}
\item {Put $\tau=s_k^2$.}
\item {Calculate $z^{(\tau)},w^{(\tau)},Z^{(\tau)},W^{(\tau)}$,
according to eqs. (\ref{zwt}-\ref{ZWt}).}
\item {Calculate the unfolded distribution $x^{(\tau)}$ and
its covariance matrix $X^{(\tau)}$, according to eqs.
(\ref{xX}-\ref{Xx}).}
\end{enumerate}
}
\end{itemize}}

The vector $x^{(\tau)}$ and matrices $X^{(\tau)}$ and $X^{-1}$ form the
complete solution of the unfolding problem defined by the matrix $A$,
simulated initial distribution $x^{\mathrm{ini}}$, the measured vector $b$
and its covariance matrix $B$.

\section{Examples}\label{Examples}

The use of the unfolding procedure described above is now illustrated
with two examples.

The first one is rather academic, and we have included it only because
the same example was used in \cite{Blobel} and \cite{Schmel}.
The response function ${\hat{\cal{A}}}(y,y^{\mathrm{true}})$ is given
by
\begin{equation}\label{blex1}
{\hat{\cal{A}}}(y,y^{\mathrm{true}})=[1-0.5(1-y^{{\mathrm{true}}\;2})]
\{4{\mathrm{exp}}[-50(y-y^{\mathrm{true}}+0.05y^{{\mathrm{true}}\;2})^2]\}.
\end{equation}
Then the probability response matrix was built according to eq. (\ref{Arat})
for 40 equidistant bins in the interval (0,2) for both $y^{\mathrm{true}}$
and $y$. The matrix is presented in Fig.~1a.
The true continuous distribution is taken to be
\begin{equation}\label{blex2}
{{\cal{X}}}(y^{\mathrm{true}})={{4}\over{4+(y^{\mathrm{true}}-0.4)^2}}
              +{{0.4}\over{0.04+(y^{\mathrm{true}}-0.8)^2}}
              +{{0.2}\over{0.04+(y^{\mathrm{true}}-1.5)^2}}.
\end{equation}
After the convolution (\ref{Axy}), the distribution ${\cal{B}}(y)$ was
discretized according to (\ref{bi}).
Simulating a counting experiment, a random normally distributed error
was then added to each entry, assuming the overall initial
statistics of 5000 events. The resulting "measured" distribution $b$ is
plotted in Fig.~1b by a dotted line. The distortions caused by the
measurement process can be seen by comparing the latter to the true
distribution (\ref{blex2}), shown by the solid curve in Fig.~1b.

The unfolding algorithm described above was then applied to the
distribution $b$. Fig.~1c shows the plot of the rescaled and rotated
r.h.s. vector $d$. The solid line corresponds to
the actual measured histogram, and
the horizontal dashed line shows the one standard deviation
statistical error in $d_i$, which is equal to 1 for each $i$. One can
see that after $i=10$ the components $d_i$ are clearly non-significant.
The flatness of this distribution for $i>10$ and its apparent compatibility
with the expected gaussian distribution with zero mean and unit
variance, is in fact a test of the gaussian random number generator used
to generate the errors in the measured histogram $b$.
If we limit ourselves to less than 10 equations, we lose some
significant information. Namely, the choice $k=1$ leaves effectively
only one equation, and the obtained "unfolded" distribution $x^{(1)}$
will be nothing else but a constant.
On the contrary, by taking more than 10
equations one includes rapidly oscillating components with
non-significant (and large) coefficients determined by the ratio $d_i/s_i$.
In this particular example taking $k=40$ would result in a distribution $x$
wildly oscillating with the amplitude of about 5000.

\begin{figure}\label{example_Blobel}
\epsfxsize16.cm
\centerline{\epsffile{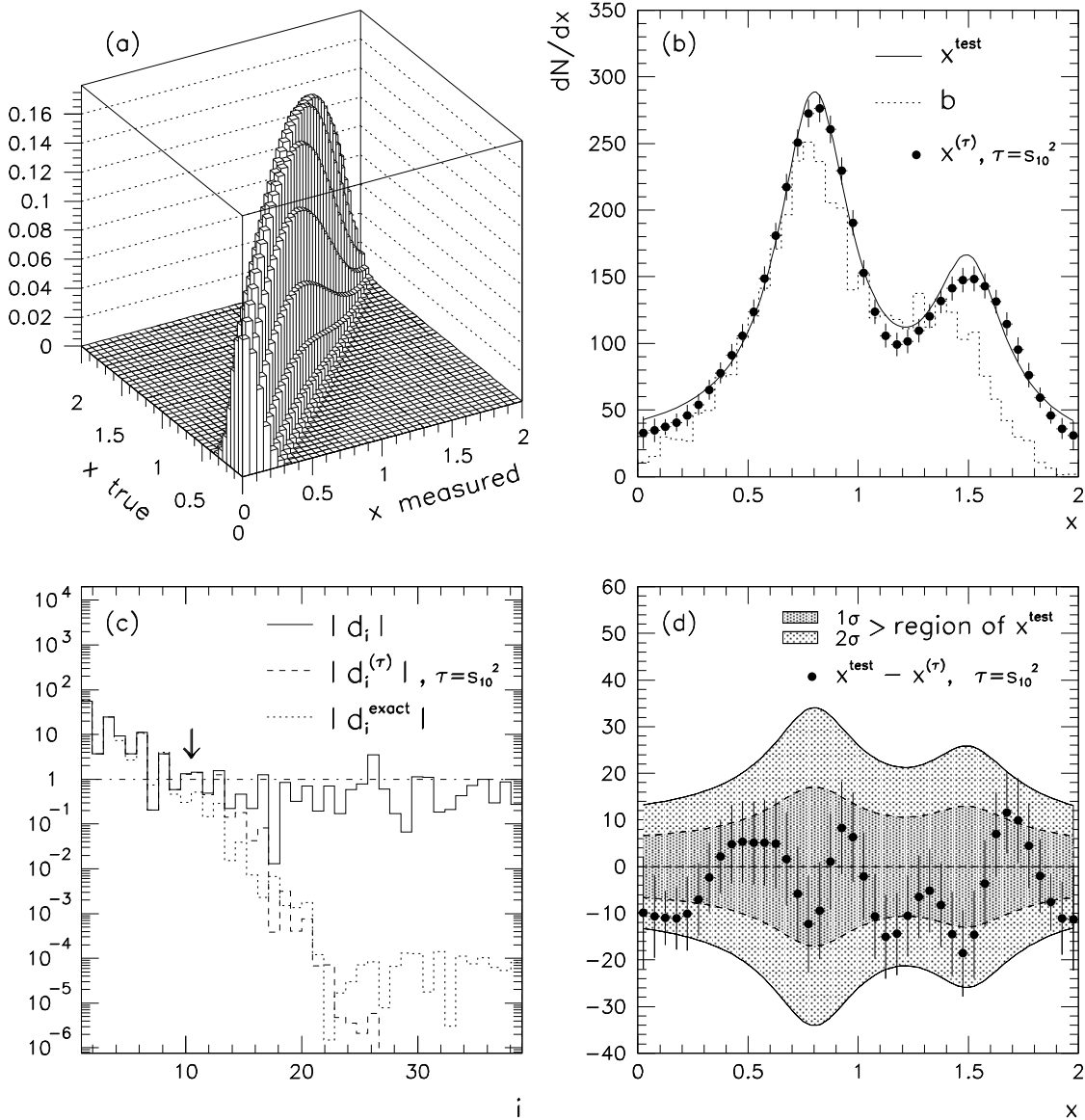}}
\caption{
{\bf{a).}} The probability matrix $\hat A$ corresponding to the
response function (\ref{blex1}).
{\bf{b).}} The true distribution ({\ref{blex2}}) (solid curve)
compared to the measured histogram $b$ and the unfolded distribution
$x^{(\tau)}$ for $\tau=s_{10}^2$.
{\bf{c).}} The absolute values of $d_i$ (solid line)
compared to the regularized r.h.s. (dashed line) and the one
unaffected by the statistical fluctuations (dotted line).
The horizontal line shows statistical errors in $d_i$, while the
arrow indicates the boundary between the significant and non-significant
equations.
{\bf{d).}} The deviation of the unfolded distribution from the true
exact one (see text for details).}
\end{figure}

The shape of the distribution $d$ suggests that the effective rank $k$
should be put equal to 10. The dashed line on Fig.~1c shows the regularized
distribution $d^{(\tau)}$ calculated using (\ref{dt}), with $\tau=s_{10}^2$.
It is interesting to compare this distribution with the similar one
calculated for the exact true distribution (\ref{blex2}) by the
same procedure of rescaling and rotation, but {\it{without}}\/
 adding the random
error (the dotted histogram in Fig.~1c). One can see that the regularized
distribution is quite close to the true exact one.

The obtained distribution $d^{(\tau)}$ is then used to
calculate the unfolded histogram
$x^{(\tau)}$, plotted in Fig.~1b (data points). It should be compared to the
true distribution (\ref{blex2}), shown by the smooth solid curve.
Note that the error bars in $x^{(\tau)}$ account only for the
diagonal elements of the covariance matrix $X$, and thus underestimate
the actual errors. The correlations between adjacent bins $x_j^{(\tau)}$  are
quite significant, so one should use the exact inverse of the covariance
matrix $X^{-1}$ for any kind of $\chi^2$ calculation
involving the unfolded vector,
and the regularized covariance matrix $X^{(\tau)}$ for the further
error propagation.

Fig.~1d presents the difference between the exact distribution and the
unfolded one, together with the bands showing one and two standard deviation
statistical fluctuations in the true exact distribution.
In this scale, one can still see
some oscillations of the unfolded solution, but they are well balanced,
distributed almost uniformly, and are confined inside the two standard
deviation band of the true solution, thus indicating that the genuine error
is about twice as large as the true statistical one would be, if the
measurements were exact.
The average $\chi^2$ over the 40 bins is equal to 0.9, meaning that
the unfolded distribution is quite satisfactory.

\begin{figure}\label{example_rho}
\epsfxsize16.cm
\centerline{\epsffile{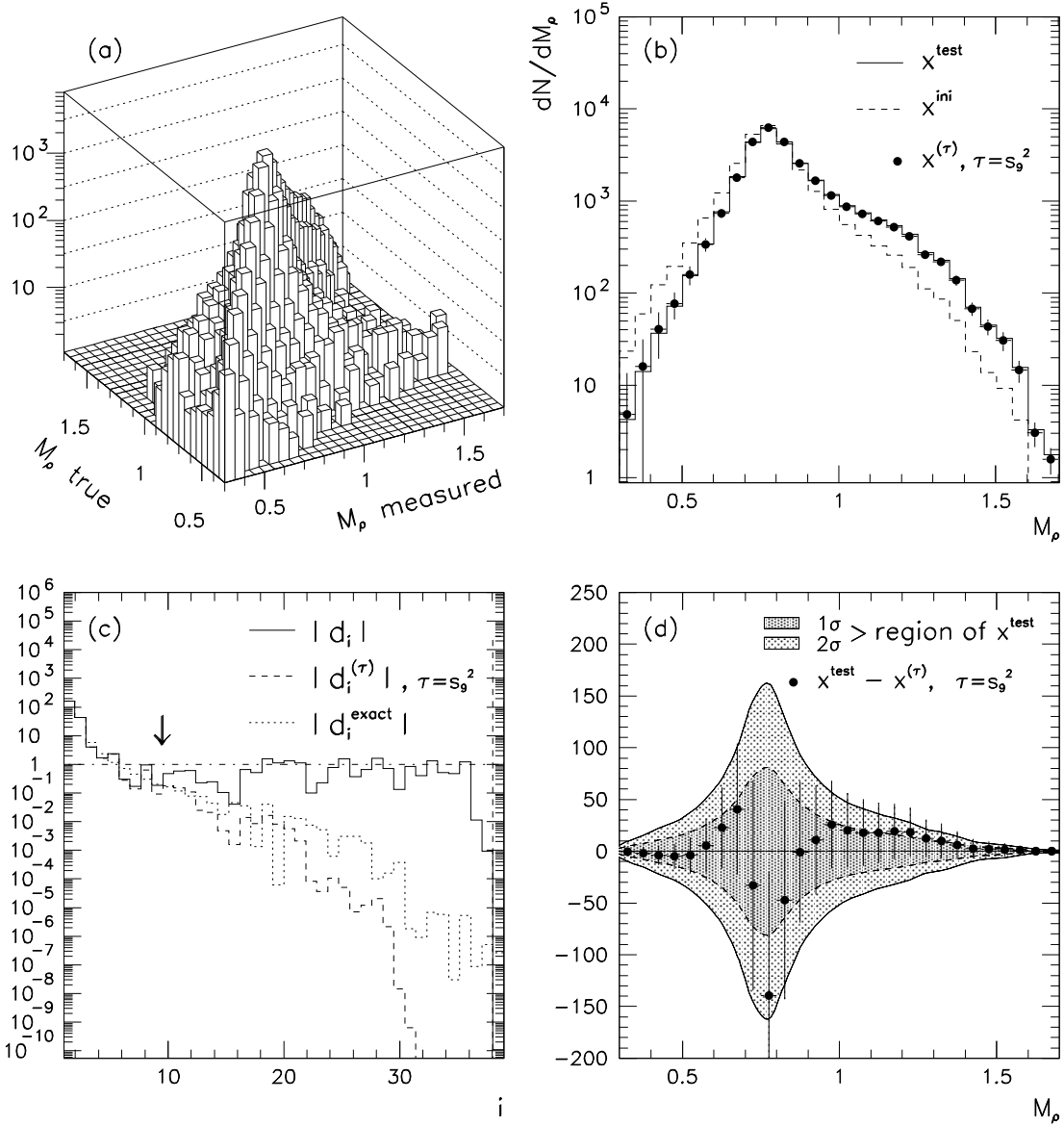}}
\caption{
{\bf{a).}} The simulated number-of-events response matrix $A$.
{\bf{b).}}
The true test distribution $x^{\mathrm{test}}$ (solid line)
compared to the unfolded one (data points). The dashed histogram
corresponds to  the initial distribution $x^{\mathrm{ini}}$
according to which the response matrix was generated.
{\bf{c).}} The absolute values of $d_i$ (solid line)
compared to the regularized r.h.s. (dashed line) and the one
unaffected by the statistical fluctuations (dotted line).
The horizontal line shows statistical errors in $d_i$, while the
arrow indicates the boundary between the significant and non-significant
equations.
{\bf{d).}} The deviation of the unfolded distribution from the true
exact one (see text for details).}
\end{figure}

In the second example we unfold a simulated spectrum of the invariant mass
of two pions, corresponding to the $\rho(770)$-meson mass region.
An artificial two-dimensional histogram reflecting a possible detector
behavior was generated as the detector response matrix $A$, and is
shown in Fig.~2a. This time it is a number-of-event matrix resulting
{}from some Monte Carlo simulation, as opposed to the probability
matrix obtained by the integration of some analytical response
function used in the previous example. The matrix is far from being
diagonal, and the initial simulated distribution $x^{\mathrm{ini}}$,
shown by a dotted line in Fig.~2b, is hardly a constant.

The "measured" distribution $b$ was obtained in a way similar to the
first example: a distribution $x^{\mathrm{test}}$ was generated
(solid line in Fig.~2b), which
has a behavior distinctly different from  $x^{\mathrm{ini}}$. The
measurement process was then simulated by the matrix multiplication:
\begin{equation}\label{mexa1}
\sum_j A_{ij}{{x_j^{\mathrm{test}}}\over{x_j^{\mathrm{ini}}}}=b_i,
\end{equation}
and finally a random gaussian error was added to each entry $b_i$,
simulating statistical fluctuations.

Rescaling and rotation results in a distribution $d_i$ plotted in Fig.~2c.
One sees that the effective rank of the system $k$ is close to 9, so the
parameter $\tau$ should
be set to the square of the 9th singular value of the matrix $AC^{-1}$.
The components $d_i$ with $i>9$ are clearly compatible with zero and have
variances close to 1, thus confirming that the errors in the measured
data are estimated correctly.

As in the first example,
the choice $k=1$ would leave us effectively with
only one equation, and the obtained "unfolded" distribution $x^{(1)}$
will be nothing else but the initial Monte Carlo distribution
$x^{\mathrm{ini}}$, shown by the dotted line in Fig.~2b.
As for the solution of the non-regularized system with $\tau=0$, it
would include all non-significant components and would oscillate
rapidly within the range $\pm(2\div3)\cdot10^4$. This solution
depends on the machine accuracy and obviously does not make any sense.

The regularized distribution $d^{(\tau)}_i$ is shown by a dashed line
in Fig.~2c. It is to be compared with the exact distribution
$d^{\mathrm{test}}$ corresponding to the vector (\ref{mexa1}) after the
same procedure of rescaling and rotation, but {\it{without}} the random
error added. One can see that the regularized vector is much closer
to the true exact one for large $i>9$.

The resulting unfolded histogram $x^{(\tau)}_i$ is shown by the data
points in Fig.~2b. The difference of the unfolded and the exact test
distributions is presented in Fig.~2d, together with one and two standard
deviation bands describing the statistical errors in the test vector.
Here, too,
the error bars show just the diagonal elements of the error matrix,
which in fact contains quite strong bin-to-bin correlations.
The agreement is very good indeed, especially if one considers the four
orders of magnitude variation range of the test distribution.

\section{Conclusion}\label{Conclusion}

The data unfolding method developed in this paper
can be used in a wide range of
applications, but is especially well-suited for high energy physics,
where the response matrix is usually estimated by a Monte Carlo
simulation of the measurement process, using some physically
motivated initial distribution of the quantity under consideration.

The extensive use of a very versatile and flexible tool --- the
Singular Value Decomposition of a matrix --- allowed us to
derive a concise loop-free algorithm for data unfolding.
Our choice of the regularization term results in smooth unfolded
distributions which have the smallest possible curvature among the
solutions satisfying the initial linear system in the least squares
sense. This is achieved by suppressing the coefficients of high-order
rapidly oscillating components of the solution.
The suppression factor
depends on both the statistical significance of the equation and
the magnitude of the corresponding singular value.
The regularized solution contains as much statistically significant
information from the measured data as possible, simultaneously
suppressing spurious, wildly oscillating components.
This suppression is a natural result of the solution process,
when the initial linear system is expanded to incorporate the minimum
curvature condition.

The method used to solve the regularized system is extremely simple and
reliable. An easy and straightforward way to  determine the
optimal value of the regularization parameter
is suggested, which allows at the same time
to test whether the quoted measurement errors are adequate.
Note that the presented solution
method is quite flexible and can be used with other choices of the
regularization term as well.

Obviously, as the number of
statistically independent data points is usually smaller
(and sometimes {\it{much}} smaller) than the number
of bins in the unfolded histogram, the latter will probably have
significant bin-to-bin correlations.
In our approach, full propagation of errors from the measured
distribution to the unfolded one is implemented, and both
the covariance matrix of the unfolded
solution and its inverse are easily calculated. This allows one to perform
further error propagation and parameter fitting without any problem,
so, contrary to the viewpoint expressed
in \cite{Blobel}, we do not think that one should use fewer bins and
custom bin boundaries for the unfolded histogram, in order to make
the covariance matrix diagonal.

Obviously, curvature minimization introduces some
systematic bias into the unfolded distribution, so the method will lead to
acceptable results only if the true solution is indeed smooth, if
the probability response matrix is used. However,  when one uses the
number-of-events response matrix, the condition of minimum curvature
means that the {\it{deviation}}\/ of the expected distribution from the
initial Monte Carlo one should be smooth enough. This clearly allows one
to use our procedure in cases when the measured distribution has some
structure and/or a wide variation range,
provided the initial Monte-Carlo has a similar behavior. If this is the case,
then even for the small
effective rank of the system, when the unfolded distribution happens to be
quite close to the initial Monte Carlo,
the former (in conjunction with the error matrix and its inverse) is still
expected to give a usable solution of the problem.

Though we have tried to present comprehensive explanations of each
step of the method, one should be able to use the algorithm formulated in
Sect. \ref{Algorithm} without having to understand in full all the details
of the underlying mathematics.
Note however, that the mathematics involved in
this paper is still much simpler than that required by other
regularization methods (e.g. {\cite{Blobel,Schmel}}) because the most
difficult tasks are successfully dealt with by the SVD procedure.

In short, the method presented in this paper allows one to unfold
data obtained by any measurement process, if the response matrix of the
detector is known. The use of the number-of-events matrix (and {\it{not}}
the probability matrix) allows one to minimize additional uncertainties
due to statistical fluctuations in the matrix itself, when the latter
is obtained using the Monte Carlo simulation process. The concise linear
algorithm results in a straightforward and simple
implementation with full error propagation,
including the complete covariance matrix and its inverse.
The method is suitable for use in a wide range of
problems, and can be generalized to incorporate multi-dimensional
distributions.

\section{Acknowledgements}

We gratefully acknowledge numerous very helpful discussions with
R.~Alemany, R.~Barlow, M.~Davier, L.~Duflot, G.Lafferty, G.Shaw
 and especially F.~Le Diberder.
 One of us (V.~K.) would
like to thank J.~Steinberger and H.~Wachsmuth for their interest and
support, and the University of Paris-Sud and LAL for their hospitality.

\end{document}